\documentstyle[12pt,aaspp4]{article}
\tightenlines
\font\boldsym=cmmib10

\def    \bOmega	{{\hbox{\boldsym\char'012}}}
\def    \bmu    {{\hbox{\boldsym\char'026}}}
\def	\be	{\begin{equation}}
\def	\ee	{\end{equation}}
\def    \ba     {\begin{eqnarray}}
\def    \ea     {\end{eqnarray}}
\def	\B	{{\rm B}}
\def	\BE	{{\rm BE}}

\def	\HH	{{\rm H}_2}

\def	\N	{{\rm N}}

\def	\ahat	{\hat{\bf a}}

\def	\dust	{{\rm dust}}
\def	\e	{{\rm e}}
\def	\el	{{\rm el}}	

\def	\bJ	{{\bf J}}

\def	\ba	{{\bf a}}

\def	\n	{{\rm n}}
\def	\nucl	{{\rm n}}	
\def	\BR	{{\rm BR}}
\def	\NR	{{\rm NR}}

\def	\cm	{\,{\rm cm}}

\def	\g	{\,{\rm g}}

\def	\Ia	{I_\parallel}

\def	\s	{\,{\rm s}}
\def    \simlt  {\lower.5ex\hbox{$\; \buildrel < \over \sim \;$}}
\def    \simgt  {\lower.5ex\hbox{$\; \buildrel > \over \sim \;$}}
\def	\gtsim	{\simgt}
\def	\ltsim	{\simlt}

\def\pmb#1{\setbox0=\hbox{#1}%
\kern-.025em\copy0\kern-\wd0	
\kern-.05em\copy0\kern-\wd0
\kern-.025em\raise.0433em\box0}

\def\plotBTD#1#2{%
  \expandafter\ifx\csname epsfbox\endcsname\relax
    \immediate\write16{%
        You need to input epsf; I'll do it for you
    }%
    \input epsf
  \fi
  \epsfysize=#2
     \openin 1 #1 \ifeof 1
        \immediate\write16{Can't open #1}%
        \vskip \the\epsfysize
      \else
         \closein 1
         \centerline{\epsfbox{#1}}%
      \fi
}

\lefthead{Lazarian \& Draine}
\righthead{Internal Relaxation}

\begin{document}

\title{Nuclear Spin Relaxation within Interstellar 
Grains\footnote{This paper is dedicated to the memory 
of E.M. Purcell.}}

\author{A. Lazarian \& B. T. Draine}
\affil{Princeton University Observatory, Peyton Hall, Princeton,
NJ 08544}

\begin{abstract}

We identify a
new mechanism of internal 
dissipation of rotational kinetic
energy in spinning dust grains,
arising from the reorientation of 
nuclear angular momentum, e.g., spins of protons.
Grain rotation induces magnetization of the nuclear spin system, 
with net alignment of nuclear spins parallel to the
grain angular velocity.
When the grain does not rotate around a principal axis,
the nuclear magnetization vector
precesses in grain body coordinates,
resulting in dissipation of
energy. 
The analogous process involving electron spins was
discovered by Purcell and termed ``Barnett relaxation''. 
We revisit the physics of the ``Barnett relaxation'' process
and correct the 
estimate for the Barnett relaxation rate.
We show
that nuclear relaxation can be orders of magnitude more important
than Barnett relaxation. This finding implies that the
processes of ``thermal flipping'' and ``thermal trapping''
are important for a broad range
of grain sizes.
\end{abstract}


\section{Introduction}

The polarization of starlight, first discovered by Hiltner (1949) and
Hall (1949), implies that the axes of grains are partially aligned with the
interstellar magnetic field. Alignment of a rotating interstellar grain,
whatever its cause may be (see a list of alignment mechanisms in
Lazarian, Goodman \& Myers 1997), requires the alignment of the grain's
body axes with its angular momentum $\bJ$.
Purcell (1979) noted that internal relaxation
of energy within a rotating grain should cause alignment of 
the grain's axis of maximal moment of inertia with $\bJ$.
In the same
pioneering study Purcell calculated the timescales for
internal relaxation and found them to be much shorter than the
gaseous damping time.
He discovered the process that he called 
``Barnett relaxation'' and identified it as 
apparently
the dominant mechanism of internal relaxation. 

Barnett relaxation is associated with the Barnett effect,
i.e., the spontaneous magnetization of a rotating paramagnetic body 
(see Landau \& Lifshitz 1962). A paramagnetic body has unpaired
electron spins and the Barnett effect
can be understood in terms of a rotating
lattice sharing its angular momentum with the 
electron
spin system.
As a result, the rotating body
becomes magnetized with magnetization $\bf M$ 
antiparallel
to the angular velocity $\bOmega$. 
For a freely rotating body $\bOmega$ and therefore
$\bf M$ precess about 
$\bf J$ in inertial coordinates; in body coordinates $\bOmega$ and $\bf M$
precess around the body's axis of maximal moment of inertia
(see Purcell 1979). This causes
time-varying magnetization of the grain material and therefore
entails dissipation.

Spitzer \& McGlynn (1979) related the Barnett relaxation time scale with
the degree of disorientation of suprathermally (i.e. much faster than
thermally) rotating grains during crossovers. The effect of suprathermal
rotation was  discovered by Purcell (1979), who identified 
recoils from H$_2$ formation events, occurring at catalytic sites on the
grain surface, as the dominant mechanism of suprathermal spin-up.\footnote{%
	A competing mechanism of spin-up due to radiation torques was 
	recently proposed by Draine \& Weingartner (1996, 1997).
	}

The theory of crossovers
was revised recently 
(Lazarian \& Draine 1997, 1999; hereafter LD97, LD99)
to include thermal fluctuations within the grain material
(Lazarian 1994, Lazarian \& Roberge 1997).
The main conclusion by 
LD97
is that 
grains with radii $a>a_c\approx 1.5\times 10^{-5}$~cm experience only
minor disalignment 
during crossovers if Barnett relaxation is 
the dominant dissipational mechanism.
As grains are essentially immune from disorientation during their phases
of suprathermal
rotation, during which paramagnetic relaxation (Davis \& Greenstein 1951)
aligns them 
with the magnetic field,
the modest disalignment
during crossovers entailed nearly perfect alignment of 
$a \gtsim 1.5\times10^{-5}\cm$
interstellar grains
in diffuse clouds
with the interstellar magnetic field. 

LD99
showed that for grains smaller than $a_c$ the
crossovers are mostly performed via 
the process of ``thermal flipping'' and this 
would entail a high degree of grain alignment if it were not
for the associated process of ``thermal trapping''; 
the latter 
occurs when thermal flipping is so rapid as to effectively average out
the systematic torques due to $\HH$ formation, etc.
We note that the value
of $a_c$ was calculated in 
LD97 and LD99
assuming 
Barnett relaxation to be the dominant mechanism of internal energy
dissipation. In the presence of more effective relaxation mechanisms
the coupling between rotational and vibrational degrees of
freedom would be larger 
thermal flipping should
be facilitated,
and the value of $a_c$ would be increased.

In this paper we show that 
the dynamics of the nuclear spins\footnote{
	For brevity, we will use the term ``nuclear spin'' to refer to the
	{\it total} nuclear angular momentum $I\hbar$.}
leads to an
important new process of internal dissipation in a rotating grain.
Although the role of nuclear spins for paramagnetic
relaxation was mentioned in the classic study by Jones \& Spitzer (1968),
and their role in developing a Barnett moment in a non-paramagnetic
substance was pointed out
by Purcell (1979), the internal relaxation 
associated with these spins was overlooked. 

Below we describe this
new magneto-mechanical effect, which we term ``nuclear relaxation''.
We show that under a broad range of conditions this process
dominates the dissipation of rotational kinetic energy in a rotating
grain.
As a result, ``nuclear relaxation'' plays an essential role in the
dynamics and alignment of interstellar grains.

\section{Dynamics of Nuclear Spins}

Purcell (1978) noted that an analog of the Barnett effect exists
for nuclear spins.\footnote{%
	This effect could be called the ``nuclear Barnett effect'' as
	opposed to the ordinary or ``electron Barnett effect''.}
If a rotating body has initially an
equal number of nuclear spins directed
parallel and anti-parallel to the angular velocity $\bOmega$, 
it can decrease its kinetic energy, at constant total angular
momentum $\bf J$, if some of the angular momentum is transfered to the
nuclear spin system. 
Increasing the projection of the nuclear angular
momentum along $\bJ$ by $+\hbar$
(at constant $J$)
reduces the rotational kinetic energy by $\hbar \Omega$.
If the rotating body is allowed to come into thermal equilibrium (without
exchanging angular momentum) with
a heat reservoir of temperature $T_\dust$ then particles of spin $S$
develop a net alignment per particle
\be
\frac{\sum_{m=-S}^Sm\exp(m\hbar\Omega/kT_\dust)}
	{\sum_{m=-S}^S\exp(m\hbar\Omega/kT_\dust)} ~.
\label{eq:tanh}
\ee
Note that this
does not depend on the magnetic moment 
$\mu$. 

In Table 1 we list a number of non-zero spin nuclei which could be
important for interstellar grains.
Carbonaceous grains will presumably contain $^{13}$C at the $\sim1\%$
level ($n\approx 10^{21}\cm^{-3}$), and $^1$H at perhaps the 10\% level
($n\approx 10^{22}\cm^{-3}$).
Silicate grains will contain $^{29}$Si at the $\sim1\%$ level
($n\approx10^{21}\cm^{-3}$), and could also contain comparable
abundances of $^{57}$Fe, $^{27}$Al, and $^{55}$Mn, as well as $^1$H.
For purposes of discussion we will assume the nuclear spin system
to consist of protons with $n_\nucl=10^{22}\cm^{-3}$.

As the number of 
parallel and antiparallel spins becomes different the body develops 
magnetization. 
The relation between $\Omega$ and the 
strength of the ``Barnett-equivalent'' magnetic
field 
$H_\BE^{\rm (n)}$
(Purcell 1979) 
that would cause the same 
nuclear
magnetization (in a nonrotating
body) is given by 
\be
{\bf H}_\BE^{(\rm n)}=\frac{\hbar}{g_{\rm n}\mu_\N}\bOmega~~~,
\label{eq:H_BE}
\ee
where $g_{\rm n}$ is the so-called nuclear $g$-factor (see Morrish 1980),
and $\mu_\N\equiv e\hbar/2m_{\rm p}c$ is the nuclear magneton,
smaller than the Bohr magneton by the electron to proton mass
ratio, $m_{\rm e}/m_{\rm p}$.

Consider a classical picture that would correspond to the
concept of the Barnett-equivalent magnetic field.
Imagine the (proton) spins as small positively charged 
tops within the rotating body.
In body coordinates their motion 
is given by
\be
\frac{d\bmu}{dt'}={\bmu\times\bOmega}~~~,
\ee
where $\bOmega$ is the angular velocity 
of the body
in an inertial frame, and
the prime signifies differentiation with respect to the rotating
frame. It is evident that this equation is similar to the equation of motion
in an inertial 
frame
if a ``Barnett-equivalent''
magnetic field defined by eq.~(\ref{eq:H_BE}) were present.
In other words, the motion of spins in the rotating
body-coordinate frame
is as though a fictitious
``Barnett-equivalent'' magnetic field 
$H_\BE^{\rm (n)}$
were present in a nonrotating body.

\section{Nuclear Susceptibility}

The zero-frequency
paramagnetic susceptibility of the nuclear spins with magnetic moments
$\mu_\nucl=g_{\rm n}\mu_\N$ is given by 
Curie's Law (Morrish 1980)
\be
\chi_\nucl(0)
=
\frac{n_\nucl\mu_\nucl^2}{3 k T}
=
4.1\times10^{-11}
\left(\!\frac{\mu_\nucl}{\mu_\N}\!\right)
^2 
\!
\left(\!\frac{n_\nucl}{10^{22}{\rm cm}^{-3}}\!\right)
\left(\!\frac{15~{\rm K}}{T}\!\right).
\label{4}
\ee
We wish to estimate the time scale on which the net alignment of 
the nuclear
spin system will respond to changes in 
the direction of
$\bOmega$.
When an oblate grain rotates freely about an axis that does not coinside
with its principal axis of largest moment of inertia $\ahat$, 
$\bOmega$ precesses
about $\bf J$. Due to the Barnett effect the 
magnetization of
nuclear spins should follow $\bOmega$. 
Interactions within the nuclear 
spin system (i.e., coupling of one
nuclear spin to the magnetic field arising from other nuclear spins)
conserve spin angular momentum, and therefore cannot change the
direction of magnetization. However, the 
component of $\bOmega$ along $\bJ$ is constant, and so even very slow
exchange of angular momentum between the lattice and the nuclear spin
system will result in magnetization of the nuclear spin system along
$\bJ$.
Since $\bJ$ precesses in body coordinates, even magnetization along $\bJ$
will have dissipation associated with it.

Consider first the coupling of the nuclear spins to the 
electron spin system.\footnote{%
	We use the term ``electron spin system'' to refer to the
	system of paramagnetic ions and electrons.
	}
The latter is coupled to the lattice on the ``spin-lattice'' timescale
$\tau_{\rm sl}$.
If we assume that the electron spin system reacts relatively quickly to 
the change in 
rotational velocity $\bOmega$, then the nuclear spin system will become
aligned by exchange of angular momentum with the electron spin system.

The ``internal'' magnetic field due to the electron spin system is
$H_i\approx 3.8 n_e\mu_e$ (van Vleck 1937), where $n_e$ is the density of
``unpaired'' electrons, and $\mu_e \approx \mu_\B$ is the magnetic moment
per unpaired electron, where $\mu_\B=e\hbar/2 m_\e c$ is the Bohr
magneton.
Because the electrons themselves precess with characteristic frequency
$\omega_e=g_e\mu_\B H_i/\hbar$,
where $g_e\approx2$ is the electron ``g-factor'',
the correlation time for $H_i$ is of order $\omega_e^{-1}$.
The fluctuating magnetic field causes the nuclear spins to undergo
a random walk; each step lasts $\sim\omega_e^{-1}$, and the change
in direction of the nuclear moment per step is $\delta\theta\approx
(g_\n\mu_\N H_i/\hbar)\omega_e^{-1}\ll 1$.
The initial phase relationship is lost after
a time $\tau_{\nucl\e}\approx
(1/\delta\theta)^2\omega_e^{-1}$. Therefore the time
for the nuclear spins to decorrelate with their 
initial direction -- which is also the time scale on which they
``relax'' to the new alignment of eq.(\ref{eq:tanh}) -- is 
\be
\tau_{\nucl\e} 
\approx
\frac{\hbar g_e}{3.8n_e g_\n^2\mu_\N^2}
\approx
3.0\times10^{-4}\left(\frac{2.7}{g_\nucl}\right)^2
\left(\frac{10^{22}\cm^{-3}}{n_e}\right)\s~.
\label{eq:tau_nucl}
\ee
The nuclear spin system can exchange angular momentum with
the electron spin system on the timescale $\tau_{\nucl\e}$.

Curiously enough a quicker nuclear spin-spin relaxation arises
from the direct interaction of nuclear moments. The magnetic field
of the neighboring nuclei is $\sim3.8n_\nucl\mu_\nucl$, and
the nuclear spin-spin relaxation time is
\be
\tau_{\nucl\nucl}\approx \frac{\hbar}{3.8g_\nucl n_\nucl \mu_\nucl^2}
=\tau_{\nucl\e} \left(\frac{g_\nucl}{g_\e}\right)\left(\frac{
\mu_\N}{\mu_\n}\right) \left(\frac{n_e}{n_{\rm n}}\right)\approx
0.58 \tau_{\nucl\e}  \left(\frac{n_e}{n_{\rm n}}\right)~.
\ee
For grains with high percentage of 
$^1$H,
$\tau_{\nucl\nucl}$ may be a factor of several smaller
than $\tau_{\nucl\e}$.
The net nuclear relaxation rate
$\tau_\nucl^{-1}=\tau_{\nucl\nucl}^{-1}+\tau_{\nucl\e}^{-1}$.

For an applied magnetic field of constant magnitude, 
rotating at frequency $\omega$,
the magnetic susceptibility contributed by the nuclear spin system
can be estimated to be (see Draine \& Lazarian 1999)
\be
\chi_\nucl(\omega)\approx \chi_\nucl(0) 
\frac{1}{(1-i\omega\tau_\nucl/2)^2}
\ee
\be
\chi_\nucl^{\prime\prime}(\omega) \equiv {\rm Im}\left[\chi_\nucl(\omega)\right]
= 
	\chi_\nucl(0)
	\frac	{\omega\tau_\nucl}
		{[1+(\omega\tau_\nucl/2)^2]^2}
\ee

\section{BARNETT RELAXATION REVISITED}

Purcell (1979) assumed that the dissipation arising from the
Barnett effect in a rotating grain is the same as would occur if 
the Barnett equivalent magnetic field were rotating in a stationary grain.
The analogy, however, is not complete. 
Consider a freely rotating 
oblate grain with eigenvalues of the moment of inertia tensor
$\Ia,I_\perp,I_\perp$, with $\Ia > I_\perp$.
Let $\ahat$ be the principal axis of largest moment of inertia, $\Ia$.
If the grain is not rotating
about a principal axis, 
then in inertial coordinates, $\ahat$ and the angular velocity
$\bOmega$ each precess around the 
In grain body coordinates,
the angular velocity $\bOmega$ precesses around $\ahat$;
the frequency
of this precession is 
\be
\omega_{1}=(h-1)\Omega\cos\theta_\Omega = 
\frac{(h-1)J\cos\theta_J}{\Ia}
\ee
(see
Purcell 1979), where $h\equiv \Ia/I_{\bot}$, $\theta_\Omega$ is the angle
between $\bOmega$ and $\ahat$, and
$\theta_J$ is the angle between $\bJ$ 
and $\ahat$.\footnote{%
	The two are related via 
	$\tan\theta_{\Omega}=h\tan\theta_{J}$.
	}
%

Let $\tau_\el\approx10^{-6}\s$ be the electron spin-lattice coupling time.
In the limit $\omega_1\tau_\el\ll 1$, 
the electron spin system will respond to the
instantaneous angular velocity, with a magnetization
$\chi H_\BE^{\rm (e)}$, where
$H_\BE^{(e)}\equiv \hbar\Omega/g_e\mu_\B$ is the Barnett-equivalent
field for the electrons.
In body coordinates, this
magnetization has a component $\chi 
H_\BE^{\rm (e)}
\sin\theta_\Omega$
which is rotating around $\ahat$, and as a result there will be
energy dissipation at a rate
$
dE/dt = \chi_e^{\prime\prime}(\omega_1)V 
(H_\BE^{\rm (e)})^2
\sin^2\theta_\Omega
$.

In the opposite limit $\omega_1\tau_\el\gg 1$,
the
electron spin system is almost decoupled from
the lattice.
The electron spins interact magnetically, but the total spin of the
electrons is conserved,
and cannot follow 
$\bOmega$ as it precesses around $\bJ$.
However, since the spin lattice coupling will not be zero,
eventually the electron spin system {\it will} become magnetized
in response to the (stationary) component of $\bOmega$ along $\bJ$,
with a magnetization 
$\chi 
H_\BE^{\rm (e)}
\cos(\theta_\Omega-\theta_J)$.
This magnetization has a component 
%
$\chi 
H_\BE^{\rm (e)}
\cos(\theta_{\Omega}-\theta_J)\sin\theta_J$ perpendicular
to $\ahat$, and in body coordinates this component of the magnetization
will rotate at frequency $\omega_1$.
We can write the energy dissipation rate as
\be
dE/dt = \chi_e^{\prime\prime}(\omega_1)V 
(H_\BE^{\rm (e)})^2
\sin^2\theta_\Omega
f(\omega_1,\theta_J)
\ee
where 
\be
f \approx 
\frac{1+(\omega_1\tau_\el)^2\sin^2\theta_J\cos^2(\theta_\Omega-\theta_J)/
\sin^2\theta_\Omega}
{1+(\omega_1\tau_\el)^2}
\ee
gives the asymptotic behavior for $\omega_1\tau_\el\ll1$ and
$\omega_1\tau_\el\gg1$.

The kinetic energy of the grain 
is $(J^2/2\Ia)[(1+(h-1)\sin^2\theta_J]$. Thus,
$dE/d\theta_J=J\omega_1\sin\theta_J$,
and 
Barnett relaxation entails
\be
\frac{d\theta_J}{dt}= -
(h-1) \frac{VJ^2}{\Ia I_\perp^2}
\left(\!\frac{\hbar}{g_n\mu_\N}\!\right)^2
\left(\!\frac{
\chi_e^{\prime\prime}
}{\omega_1}\!\right)
\sin\theta_J\cos\theta_J f(h, \theta_J),
\ee
The Barnett relaxation is slower than estimated by Purcell (1979) by 
the
factor $f < 1$.
However, this factor is of order unity, and we will take $f=1$ in the
remaining discussion.

%
For an $a\!\times\!a\sqrt{3}\!\times\!a\sqrt{3}$ 
brick, Barnett
relaxation gives
\be
\tau_\BR
\approx2.1\times10^8 \hat{\rho}^2 a_{-5}^7 \left(\frac{J_d}{J}\right)^2
\left[1+\left(\frac{\omega_1\tau_{\rm el}}{2}\right)^2\right]^2
\s 
\ee
where $\hat{\rho}\equiv\rho/(2\g\cm^{-3})$,
$\hat{a}\equiv a/10^{-5}\cm$
and
$J_d\equiv[\Ia kT_d/(h-1)]^{1/2}$
is a characteristic angular momentum appearing in the theory
of crossovers 
(LD97, LD99).

\bigskip\plotBTD{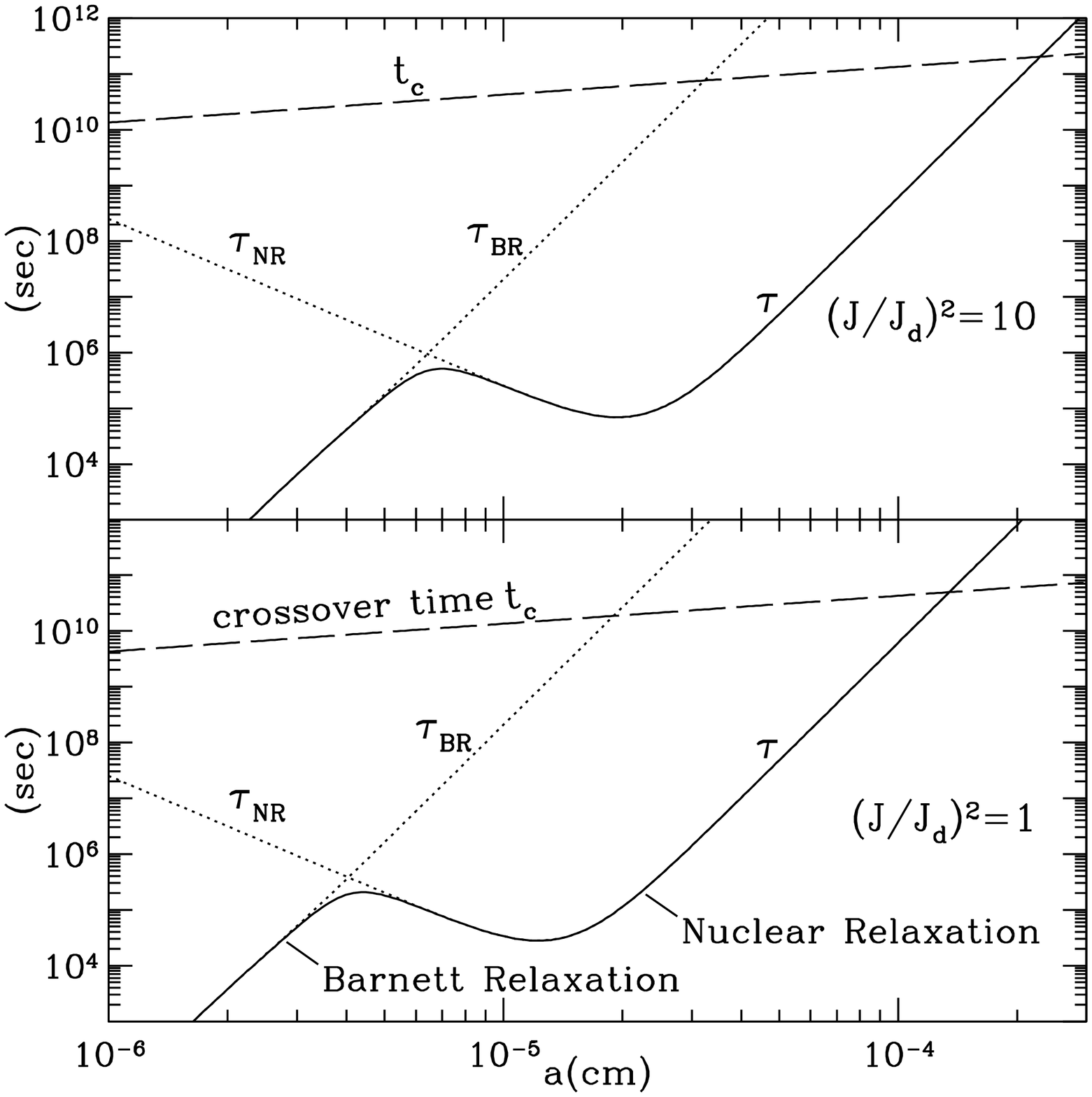}{9.0cm}
\figcaption[f1.eps]{
	Time $\tau$ for alignment of $\ahat$ with $\bJ$
	for a freely rotating
	grain, due to dissipation associated with nuclear spins
	(nuclear relaxation) and electron spins
	(Barnett effect).
	Results are shown for grains with $(J/J_d)^2=1$ (lower panel)
	and 10 (upper panel).
	Also shown is the ``crossover time'' $t_c = J/\dot{J}_\parallel$, 
	where the torque $\dot{J}_\parallel$ is due to H$_2$ formation
	with a density of active sites $10^{13}\cm^{-2}\cm^{-2}$.
	The results shown here are for an $a\!\times\!b\!\times\!b$
	``brick'', with $b/a=\sqrt{3}$, spinning with $\cos\theta_J=0.5$.
	The grain is assumed to have
	$\rho=2\g\cm^{-3}$, 
	unpaired electrons with $n_e=10^{22}\cm^{-3}$,
	and protons with $n_\n=10^{22}\cm^{-3}$.
	\label{fig:tau_B,tau_NR}
        }

\section{Nuclear Relaxation In a Rotating Grain}

Nuclear relaxation is more 
effective than
Barnett 
relaxation.  The
time for nuclear relaxation to 
align $\ahat$ 
with $\bJ$ is
\be
\tau_\NR 
=
\!\frac{I_a I_\perp^2}{(h-1)VJ^2}
\frac{g_\nucl^2\mu_\N^2}{\hbar^2}
\frac{1}{\chi_\nucl(0)\tau_\nucl}
\left[1\!+\!\left(\frac{\omega_1\tau_\nucl}{2}\right)^2\right]^2
\ee
%
For
an $a\!\times\!b\!\times\!b$ brick
with $b/a=\sqrt{3}$, we have $h-1=1/2$, and
\be
\tau_\NR
\approx 610
\hat{\rho}^2
a_{-5}^7
\!
\left(\!\frac{J_d}{J}\!\right)^2
\!
\left(\!\frac{n_e}{n_\nucl}\!\right)
\!
\left(\!\frac{g_\nucl}{3.1}\!\right)^4
\!
\left(\!\frac{2.7\mu_\N}{\mu_\nucl}\!\right)^2
\!
\left[1\!+\!\left(\!\frac{\omega_1\tau_\nucl}{2}\!\right)^2\right]^2{\rm s}
\label{eq:tau_NR}
\ee

Figure 1 shows 
$\tau_\NR$, $\tau_\BR$ and the overall relaxation time
$\tau=(\tau_\NR^{-1}+\tau_\BR^{-1})^{-1}$
as a function of grain size,
for grains rotating with 
$J/J_d=1$ and $J=10^{1/2}J_d$.
For slow rotation rates (large $a$)
the timescale $\tau_\NR$ is $\sim10^6$ times
shorter than $\tau_\BR$!
As a result
the critical grain size for which thermal fluctuations dominate the dynamics
of crossovers is increased by an order of magnitude. 
We conclude that ``thermal flipping'' 
(LD99)
must therefore be an
essential part of crossover dynamics for all
grains with $a\ltsim10^{-4}\cm$.

%
\section{Discussion}	

The first striking question is why an effect so feeble as nuclear magnetism
can be so important in terms of internal relaxation. 
The answer 
is that spins align in a rotating body because of their
angular momentum, not because of their magnetic moments.
The magnetic moment enters only as a means for the spins to exchange
angular momentum with the lattice.
The coupling 
%
within the
electron spin system 
%
is very effective,
with the result that there is minimal ``lag'' in the electron spin
alignment when the grain angular velocity $\bOmega$ changes in grain
body coordinates.
In the case of the nuclear spin system, however, the value of the nuclear
magnetic moment is large enough to provide significant
%
spin-spin
 coupling, 
but 
weak enough so that there is a significant
lag in the nuclear spin alignment when $\bOmega$ precesses around the
grain axis $\ahat$: the coupling is ``just right'' for nuclear relaxation
to be extremely effective for $10^{-5}\ltsim a \ltsim 10^{-4}\cm$ 
grains.
The efficiency of nuclear relaxation drops for sufficiently high
frequencies and therefore
Barnett relaxation dominates for $a\ltsim5\times10^{-6}\cm$,
as seen in Figure \ref{fig:tau_B,tau_NR}.

LD99
found that ``thermal flipping'' was an
important 
element
of grain dynamics for sufficiently small grains.
Assuming the Barnett effect to dominate internal dissipation,
they
estimated that thermal flipping would
be important for $a\ltsim 10^{-5}\cm$ grains, and they speculated that
for $a\ltsim 5\times10^{-6}\cm$ grains, thermal flipping would be so
rapid as to interfere with the ability of systematic torques to spin
these grains up to suprathermal rotation rates.

In the present paper we have shown that internal relaxation associated
with nuclear spin alignment
can be many orders of magnitude more rapid than due to the Barnett effect
for $a\gtsim 10^{-5}\cm$ grains.
As a result, the phenomena of thermal flipping and thermal trapping 
become important for grains as large as $\sim\!10^{-4}\cm$.

\section{Conclusions}

The results of this study can be summarized as follows:

1. Freely rotating interstellar grains are subject to ``nuclear
relaxation'' when their angular momentum $\bJ$ is not along a
principal axis.

2. For slowly rotating interstellar grains the dissipation due to
nuclear relaxation is many orders of magnitude higher than
that due to Barnett relaxation.

3. The critical size 
$a_c$ below which thermal fluctuations are 
effective during a crossover increases to $\sim\!10^{-4}$~cm.
Thermal flipping and thermal trapping must therefore be essential
elements of the dynamics of all $a\ltsim10^{-4}\cm$ interstellar grains.

\acknowledgements
We thank R.H. Lupton for availability of SM.
This research was supported in part by NASA grant NAG5-7030 and 
NSF grant AST-9619429.


\begin{center}
\begin{deluxetable}{c c c c c}
\tablecolumns{5}
\tablewidth{0pc}
\tablecaption{Nuclear Properties\tablenotemark{a}\label{tab:spins}}
\tablehead{
\colhead{Nucleus}&
\colhead{$I$\tablenotemark{b}}&
\colhead{Abund.(\%)}&
\colhead{$\mu_\nucl/\mu_N$\tablenotemark{c}}&
\colhead{$g_\n$}
	}
\startdata
$^1$H&		1/2&	99.998&	2.675&	3.089\\

$^{13}$C&	1/2&	1.108&	0.673&	0.777\\

$^{14}$N	&1	&99.635	&0.193	&0.137\\  

$^{17}$O	&5/2	&0.037	&0.363	&-0.123\\

$^{35}$Cl	&3/2	&75.53	&0.262	&0.135\\

$^{27}$Al	&5/2	&100	&0.697	&0.236\\

$^{31}$P	&1/2	&100	&1.083	&1.250\\

$^{57}$Fe	&1/2	&2.19	&0.086	&0.100\\

$^{29}$Si	&1/2	&4.70	&0.531	&-0.614\\

$^{55}$Mn	&5/2	&100	&0.660	&0.223\\
\enddata
\tablenotetext{a}{\footnotesize From Robinson (1991)}
\tablenotetext{b}{\footnotesize total angular momentum quantum number}
\tablenotetext{c}{\footnotesize%
	$\mu_\N\equiv e\hbar/2m_p c=5.05\times 10^{-24}$~erg/gauss.}
\end{deluxetable}
\end{center}

\end{document}